\documentclass[aps,showpacs,preprintnumbers,amsmath, amssymb]{revtex4}

\usepackage{float}
\usepackage{graphics,epsfig}
\usepackage{graphicx}
\usepackage{dcolumn}
\usepackage{bm}

\begin{document}
\baselineskip=0.8 cm

\title{{\bf No hair theorem for spherically symmetric regular compact stars with Dirichlet boundary conditions}}
\author{Yan Peng$^{1}$\footnote{yanpengphy@163.com}}
\affiliation{\\$^{1}$ School of Mathematical Sciences, Qufu Normal University, Qufu, Shandong 273165, China}

\vspace*{0.2cm}
\begin{abstract}
\baselineskip=0.6 cm
\begin{center}
{\bf Abstract}
\end{center}

We study scalar condensation in the background of asymptotically flat spherically symmetric
regular Dirichlet stars. We assume that the scalar field decreases as the star
surface is approached. Under these circumstances, we prove a no hair theorem for
neutral regular compact stars. We also extend the discussion to charged regular compact
stars and find an upper bound for the charged star radius. Above the upper bound, the scalar
hair cannot exist. Below the upper bound, we numerically obtain solutions
of scalar hairy charged stars.

\end{abstract}

\maketitle
\newpage
\vspace*{0.2cm}

\section{Introduction}

Black hole no hair theorems, see e.g. \cite{Bekenstein,Chase,C. Teitelboim,Ruffini-1},
establish a most famous property in the general relativity theory. If generically true, such
no-hair property would signify that asymptotically flat black holes
are simply determined by three conserved charges, which are the mass,
angular momentum and charge of black holes, for recent progress see references \cite{Hod-1}-\cite{Ca} and reviews
\cite{Bekenstein-1,CAR}. The belief that such property generically holds is partly based
on the existence of black hole horizons, which can absorb matter and
radiation fields. Whether such no hair theorems are unique to black hole
spacetimes is an interesting question to be answered.

In fact, no hair behaviors are not restricted to the spacetime with a horizon.
In the asymptotically flat gravity, it was firstly proved in \cite{Hod-6} that
static massive scalar field hair cannot exist outside neutral horizonless
reflecting stars. With scalar fields nonminimally coupled to
gravity, no hair property also appears in the background of
asymptotically flat neutral horizonless reflecting stars \cite{Hod-7}.
And in the asymptotically dS spacetime, neutral horizonless reflecting stars
cannot support vector and tensor fields \cite{Bhattacharjee}.
In all, no hair behavior is a general property
for neutral horizonless reflecting stars.

Along this line, it is interesting to examine whether
there is no hair theorem in the charged object background.
When the charged horizonless reflecting shell radius is above an upper bound,
static scalar field hairs cannot exist and below the upper bound,
scalar fields can condense outside the shell \cite{Hod-8,Hod-9,Yan Peng-1}.
We should point out that this property is very different from cases in black holes,
where black hole no hair theorem holds for any size of the horizon \cite{bm-1,bm-2}.
Similar to the charged shell case, it was found that
large horizonless reflecting stars tend to have no scalar field hair
\cite{Hod-10,Yan Peng-2,Yan Peng-4,Yan Peng-5,Yan Peng-6,LR}.
Moreover, large horizonless stars with Neumann boundary conditions
also cannot support the static scalar field hair \cite{Yan Peng-3,YPN}.
In the present paper, we plan to extend discussions to
Dirichlet boundary conditions.

In the following, we introduce the model of
static scalar fields in the background of
horizonless stars with Dirichlet boundary conditions.
We analytically study scalar field condensations outside
neutral stars and charged stars. With numerical methods, we obtain
solutions of scalar hairy charged horizonless stars.
We summarize main results in the last section.

\section{The no scalar hair behaviors in neutral regular stars}

We study static scalar field condensations outside
asymptotically flat regular stars.
In Schwarzschild coordinates, the ansatz of the four dimensional
spherically symmetric star is \cite{mr1,mr2}
\begin{eqnarray}\label{AdSBH}
ds^{2}&=&-g(r)e^{-\chi(r)}dt^{2}+\frac{dr^{2}}{g(r)}+r^{2}(d\theta^2+sin^{2}\theta d\phi^{2}).
\end{eqnarray}
The metric solutions $\chi(r)$ and $g(r)$ depend on
the radial coordinate r. Asymptotic flatness of the spacetime requires
$\chi(r)\rightarrow 0$ and $g(r)\rightarrow 1$
as $r\rightarrow \infty$.

And the Lagrangian density is given by \cite{bl0,bl1,YPBW}
\begin{eqnarray}\label{lagrange-1}
\mathcal{L}=R-|\nabla_{\alpha} \psi|^{2}+\mu^{2}\psi^{2}.
\end{eqnarray}
Here R is Ricci curvature and $\psi$ is the scalar field with mass $\mu$.

For simplicity, we take the scalar field only depending on the radial
coordinate in the form $\psi=\psi(r)$.
Then the scalar field equation reads \cite{Hod-6}
\begin{eqnarray}\label{BHg}
\psi''+(\frac{2}{r}-\frac{\chi'}{2}+\frac{g'}{g})\psi'-\frac{\mu^2}{g}\psi=0.
\end{eqnarray}

The reflecting boundary condition can be
interpreted as a infinity repulsive potential
to suppress scalar field condensations.
And for finite potential, the scalar field
may decrease to be a nonzero constant at the boundary.
When $\psi(r_{s})=0$, it returns to the reflecting case and
no scalar field hair theorem holds \cite{Hod-6}.
It therefore remains to prove no hair theorem
for scalar fields with the behavior $\psi(r_{s})\neq0$.
In this work, we take the particular Dirichlet boundary condition that
scalar fields decrease as approaching the star surface.
At the surface, the scalar field satisfies
\begin{eqnarray}\label{InfBH}
\psi(r_{s})> 0,~~~\psi'(r_{s})\geqslant 0~~~~or~~~~\psi(r_{s})< 0,~~~\psi'(r_{s})\leqslant 0.
\end{eqnarray}

In the large r region, the scalar field asymptotically behaves
as $\psi\sim A\cdot\frac{1}{r}e^{-\mu r}+B\cdot\frac{1}{r}e^{\mu r}$,
with A and B as integral constants.
The scalar field energy density is given by
\begin{eqnarray}\label{InfBH}
\rho=-T_{t}^{t}=\frac{1}{2}[g(\psi')^2+V(\psi^2)].
\end{eqnarray}
An spacetime with finite ADM mass is characterized by \cite{Hod-6}
\begin{eqnarray}\label{InfBH}
r^3\rho\rightarrow 0~~~for~~~r\rightarrow\infty.
\end{eqnarray}
In order to obtain the physical solution with finite ADM mass, we set $B=0$.
Then the scalar field satisfies the infinity boundary condition
\begin{eqnarray}\label{InfBH}
&&\psi(\infty)=0.
\end{eqnarray}

In the case of $\psi(r_{s})> 0$ and $\psi'(r_{s})\geqslant 0$,
the function $\psi$ with $\psi(\infty)=0$ must possesses at least
one extremum point $r=r_{peak}$ above the star surface $r_{s}$.
At this extremum point, there are following relations
\begin{eqnarray}\label{InfBH}
\{ \psi>0,~~~~\psi'=0~~~~and~~~~\psi''\leqslant0\}~~~~for~~~~r=r_{peak}.
\end{eqnarray}

According to (8), we obtain the inequality at $r=r_{peak}$ in the form
\begin{eqnarray}\label{BHg}
\psi''+(\frac{2}{r}-\frac{\chi'}{2}+\frac{g'}{g})\psi'-\frac{\mu^2}{g}\psi<0.
\end{eqnarray}

In another case of $\psi(r_{s})< 0$ and $\psi'(r_{s})\leqslant 0$,
the function $\psi$ must possesses at least
one extremum point $r=\tilde{r}_{peak}$ above the star surface $r_{s}$.
At this extremum point, there is the following characteristic relation
\begin{eqnarray}\label{InfBH}
\{ \psi<0,~~~~\psi'=0~~~~and~~~~\psi''\geqslant 0\}~~~~for~~~~r=\tilde{r}_{peak}.
\end{eqnarray}

With relations (10), we obtain the following inequality at $r=\tilde{r}_{peak}$
\begin{eqnarray}\label{BHg}
\psi''+(\frac{2}{r}-\frac{\chi'}{2}+\frac{g'}{g})\psi'-\frac{\mu^2}{g}\psi>0.
\end{eqnarray}

We can see that relations (9) and (11) are in contradiction with the
scalar field equation (3). So no hair theorem holds for
asymptotically flat neutral horizonless star
with Dirichlet boundary conditions (4).

\section{The upper bound on radii of hairy charged stars}

Now we turn to study scalar condensations outside charged regular stars.
The Lagrange density describing scalar fields coupled to Maxwell fields reads \cite{bl0,bl1,YPBW}
\begin{eqnarray}\label{lagrange-1}
\mathcal{L}=-F^{MN}F_{MN}-|\nabla_{\alpha} \psi-q A_{\alpha}\psi|^{2}-\mu^{2}\psi^{2}.
\end{eqnarray}
Here $\psi(r)$ stands for the scalar field with mass $\mu$ and charge q.
$A_{\alpha}$ represents the ordinary Maxwell field.

Neglecting scalar fields' backreaction on the charged star spacetime,
we obtain solutions $\chi=0$ and $g=1-\frac{2M}{r}+\frac{Q^2}{r^2}$,
where $M$ is the star mass and $Q$ is the star charge.
The Maxwell field can be expressed as $A_{\mu}=-\frac{Q}{r}dt$.
And the scalar field equation is \cite{Hod-10,Yan Peng-2,bl2,bl3,bl4,bl5,bl6}
\begin{eqnarray}\label{BHg}
\psi''+(\frac{2}{r}+\frac{g'}{g})\psi'+(\frac{q^2Q^2}{r^2g^2}-\frac{\mu^2}{g})\psi=0,
\end{eqnarray}
where $g=1-\frac{2M}{r}+\frac{Q^2}{r^2}$.
The boundary conditions are still expressed as (4) and (7).

Introducing a new function $\tilde{\psi}=\sqrt{r}\psi$,
we can transfer the scalar field equation into
\begin{eqnarray}\label{BHg}
r^2\tilde{\psi}''+(r+\frac{r^2g'}{g})\tilde{\psi}'+(-\frac{1}{4}-\frac{rg'}{2g}+\frac{q^2Q^2}{g^2}-\frac{\mu^2r^2}{g})\tilde{\psi}=0
\end{eqnarray}
with $g=1-\frac{2M}{r}+\frac{Q^2}{r^2}$.

According to boundary conditions (4) and
$\tilde{\psi}'=(\sqrt{r}\psi)'=\frac{1}{2\sqrt{r}}\psi+\sqrt{r}\psi$, one deduces that
\begin{eqnarray}\label{InfBH}
\tilde{\psi}(r_{s})> 0,~~~\tilde{\psi}'(r_{s})\geqslant 0~~~~or~~~~\tilde{\psi}(r_{s})< 0,~~~\tilde{\psi}'(r_{s})\leqslant 0.
\end{eqnarray}

As the scalar field behaves as $\psi\sim A\cdot\frac{1}{r}e^{-\mu r}$ for large radius,
the infinity boundary condition is \cite{LR}
\begin{eqnarray}\label{InfBH}
\tilde{\psi}(\infty)=0.
\end{eqnarray}

According to (15) and (16), the function $\tilde{\psi}$ must have one extremum point $r=\bar{r}_{peak}$.
At this extremum point, there are characteristic relations
\begin{eqnarray}\label{InfBH}
\{ \tilde{\psi}^2>0,~~~~\tilde{\psi}\tilde{\psi}'=0~~~~and~~~~\tilde{\psi}\tilde{\psi}''\leqslant0\}~~~~for~~~~r=\bar{r}_{peak}.
\end{eqnarray}

With relations (14) and (17), we obtain the inequality
\begin{eqnarray}\label{BHg}
-\frac{1}{4}-\frac{rg'}{2g}+\frac{q^2Q^2}{g^2}-\frac{\mu^2r^2}{g}\geqslant0~~~for~~~r=\bar{r}_{peak}.
\end{eqnarray}

It can be transferred into
\begin{eqnarray}\label{BHg}
\mu^2r^2g\leqslant q^2Q^2-\frac{rgg'}{2}-\frac{1}{4}g^2~~~for~~~r=\bar{r}_{peak}.
\end{eqnarray}

Since $r\geqslant r_{s}> M+\sqrt{M^2-Q^2}\geqslant M \geqslant Q$, there are relations
\begin{eqnarray}\label{BHg}
g=1-\frac{2M}{r}+\frac{Q^2}{r^2}=\frac{1}{r^2}(r^2-2Mr+Q^2)=\frac{1}{r^2}[(r-M)^2-(M^2-Q^2)]\geqslant0,
\end{eqnarray}
\begin{eqnarray}\label{BHg}
rg'=r(1-\frac{2M}{r}+\frac{Q^2}{r^2})'=r(\frac{2M}{r^2}-\frac{2Q^2}{r^3})=\frac{2M}{r}(1-\frac{Q}{r}\frac{Q}{M})\geqslant 0
\end{eqnarray}
and
\begin{eqnarray}\label{BHg}
(r^2g)'=(r^2-2Mr+Q^2)'=2(r-M)\geqslant 0.
\end{eqnarray}

From (19-22), we get the relation
\begin{eqnarray}\label{BHg}
\mu^2r_{s}^2g(r_{s})\leqslant \mu^2r^2g(r)\leqslant q^2Q^2-\frac{rgg'}{2}-\frac{1}{4}g^2\leqslant q^2Q^2~~~for~~~r=\bar{r}_{peak}.
\end{eqnarray}

It yields the inequality
\begin{eqnarray}\label{BHg}
\mu^2r_{s}^2g(r_{s})\leqslant  q^2Q^2.
\end{eqnarray}

The inequality (24) can be expressed as
\begin{eqnarray}\label{BHg}
\mu^2r_{s}^2(1-\frac{2M}{r_{s}}+\frac{Q^2}{r_{s}^2})\leqslant  q^2Q^2.
\end{eqnarray}

Then we further transfer (25) into
\begin{eqnarray}\label{BHg}
(\mu r_{s})^2-(2\mu M)(\mu r_{s})+Q^2(\mu^2-q^2)\leqslant 0.
\end{eqnarray}

According to (26), we obtain bounds on hairy star radii as
\begin{eqnarray}\label{BHg}
\mu r_{s}\leqslant \mu M+\sqrt{\mu^{2}(M^2-Q^2)+q^2Q^2}.
\end{eqnarray}
When the star radius is above the upper bound (27),
scalar fields cannot exist outside the charged star.
In the following, we will show that the scalar hair
may form when the star radius is below this upper bound.

\section{Scalar field configurations supported by charged stars}

In this part, we try to obtain scalar hairy star solutions with numerical methods.
Firstly, we set $\mu=1$ with the symmetry of equation (13) expressed as
\begin{eqnarray}\label{BHg}
r\rightarrow k r,~~~~ \mu\rightarrow \mu/k,~~~~ M\rightarrow k M,~~~~ Q\rightarrow k Q,~~~~ q\rightarrow q/k.
\end{eqnarray}
Then we follow approaches in \cite{Yan Peng-2} to
search for hairy star solutions by integrating
the equation (13) from $r_{s}$ to the infinity.
We show numerical hairy compact star solutions
with $q=2$, $Q=4$ and $M=5$ in Fig. 1.

\begin{figure}[h]
\includegraphics[width=350pt]{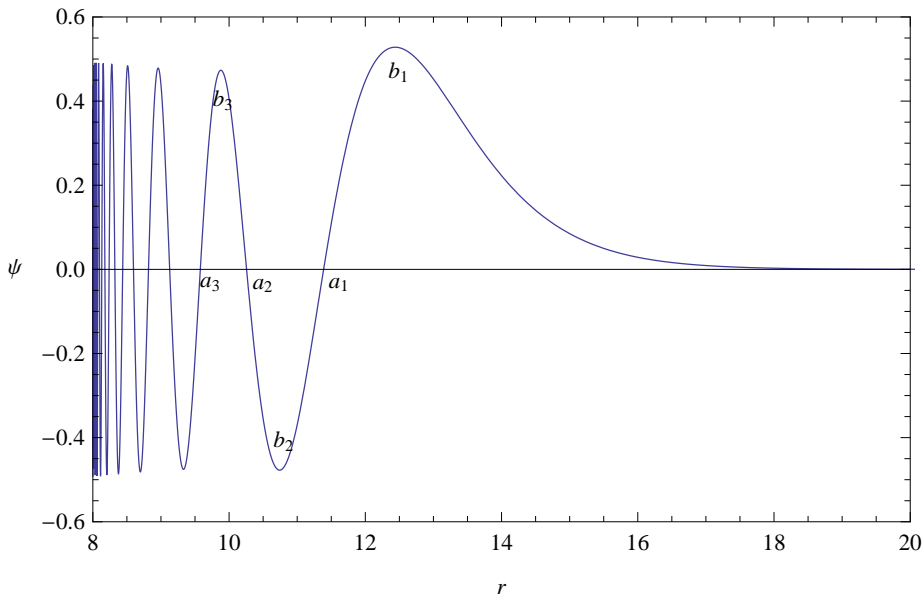}\
\caption{\label{EEntropySoliton} (Color online) We plot the function of $\psi$
with respect to the coordinate $r$ with $q=2$, $Q=4$, $M=5$.
$a_{i}$ are zero points and $b_{i}$ are extremum points.}
\end{figure}

We label zero points of $\psi(r)$ from right to left as $a_{1},~a_{2},~a_{3},\ldots$ with
$a_{i}$ corresponding to reflecting star radii in \cite{Yan Peng-2}.
We also define extremum points from right to left as $b_{1},~b_{2},~b_{3},\ldots$,
where $b_{i}$ are Neumann star radii in \cite{Yan Peng-3}.
With particular Dirichlet boundary conditions (4), the surface radius can be fixed at any point in
ranges $[a_{i},b_{i}]$. Here the largest radius is obtained at $b_{1}\thickapprox 12.440$
below the upper bound $r_{s}\leqslant 5+\sqrt{73}\thickapprox 13.544$ according to (27).

With particular Dirichlet boundary conditions (4),
we analytically prove that scalar hairy charged star radii have an upper bound (27),
which is also supported by numerical results in
the front paragraph. If we take other Dirichlet boundary
condition different from (4), such as $\psi(r_{s})>0$ and $\psi'(r_{s})<0$,
scalar hairy charged star radii can be
imposed at any point in the range $(b_{1},\infty)$.
So there is no upper bound for scalar hairy charged star radii
in the case of $\psi(r_{s})>0$ and $\psi'(r_{s})<0$.

\section{Conclusions}

We studied the gravity model of static massive scalar fields
in the background of asymptotically flat spherically symmetric
regular stars. We imposed particular Dirichlet boundary conditions that
scalar fields decrease as the star surface is approached.
Under these circumstances, we found that the Dirichlet 
neutral compact star cannot support the
existence of static massive scalar field hairs.
We also extended discussions to the charged compact star spacetime.
In the case of charged stars, we obtained an upper bound on
the star radius. Above the bound, no scalar hair
theorem holds and below the bound, we numerically
obtained solutions of scalar hairy charged stars.

\begin{acknowledgments}

We would like to thank the anonymous referee for the constructive suggestions to improve the manuscript.
This work was supported by the Shandong Provincial Natural Science Foundation of China under Grant
No. ZR2018QA008.

\end{acknowledgments}

\end{document}